\documentclass[aps,twocolumn]{revtex4}

\usepackage{graphicx,subfigure}
\usepackage{amssymb,amsmath} 
\usepackage[utf8]{inputenc}
\usepackage[english]{babel}

\usepackage{times}
\usepackage{lipsum,verbatim}

\usepackage[normalem]{ulem}
\usepackage{hyperref}

\usepackage{color}
\definecolor{red}{rgb}{1,0,0}
\definecolor{green}{rgb}{0,1,0}
\definecolor{blue}{rgb}{0,0,1}

\newenvironment{menor}[1]{\begin{list}{}{
         \leftmargin=#1  \rightmargin=#1  \parsep=0pt
         \partopsep=0pt}\item[]}{\end{list}}

\begin{document}

\title{Zip and velcro bifurcations in competition models in ecology
       and economics}

\author{Jason A.C.~\surname{Gallas}}
\affiliation{Max-Planck-Institut f\"ur Physik komplexer Systeme,
  D-01187 Dresden, Germany}
\affiliation{Complexity Sciences Center, 9225 Collins Avenue Suite 1208,  
Surfside, FL 33154, USA}
\affiliation{Instituto de Altos Estudos, 
Rua Silvino Lopes 419-2502,  58039-190 Jo\~ao Pessoa, Brazil}
\date{September 10, 2019}

\begin{abstract}
During the last six years or so, a number of interesting papers
discussed systems with line segments of equilibria,  
planes of equilibria, and with more general equilibrium configurations.
This note draws attention to the fact that such equilibria were considered
previously by Mikl\'os Farkas (1932-2007), in papers published in 1984-2005. 
He called zip bifurcations those involving line segments of equilibria,
and velcro bifurcations those involving planes of equilibria.
We briefly describe prototypical situations involving zip and velcro
bifurcations.\\

\end{abstract}

\maketitle

\section{Introduction}

About ten years ago it was realized that, in addition to the familiar chaotic
attractors associated with saddle points, dynamical systems may also contain
attractors not connected to such points, the so-called {\sl hidden}
attractors \cite{lkv11,pr16}.
This startling finding induced considerable interest and a large number of
publications concerning the properties of systems displaying no equilibria,
chaotic systems with lines of equilibria, as well as systems containing
planes and more general surfaces of equilibria.

For instance, by performing a systematic computer search among certain
families polynomial dynamical systems,
Jafari and Sprott \cite{js13} found nine chaotic flows with quadratic
nonlinearities which have the unusual feature of displaying a line segment
of equilibrium points.
Other similar polynomial flows were also investigated \cite{ls,lst}.
As remarked by these authors,
such systems belong to a newly introduced category of chaotic
systems with hidden attractors that are important and potentially problematic
in real-life applications. Wang and Chen  \cite{wc13} reported on how to
construct systems having any number of equilibria.
Uyaro\v glu and Kocamaz  \cite{uk15} investigated the control of a chaotic system
having a line of equilibria using a  passive  control  method.
A chaotic flow with a plane of equilibria was investigated by
Jafari et al.~\cite{jsm16}.
Very recently, Wu et al.~\cite{wzm19} studied a neural memristor with infinite
or without equilibrium,
while Pham and coworkers presented a gallery of chaotic systems with an infinite
number of equilibrium points \cite{p16} and
reported simulations and an experimental implementation of a system with a
line of equilibria but no linear term \cite{p19}.

The purpose of this note is to draw attention to the fact that systems
with lines and surfaces of equilibria were considered previously
by Mikl\'os Farkas (1932-2007), in a series of papers published during
the years 1984-2005.
Farkas called {\sl zip bifurcations} those involving line segments of equilibria,
and {\sl velcro bifurcations} those in systems with planes of equilibria.
In  the remainder of this note, we briefly review these concepts.

\section{Zip bifurcations}

Farkas introduced the concept of zip bifurcations in 1984 \cite{zip,f87}.
An extended and detailed presentation is given in his 1994 book
{\sl Periodic Motions} \cite{fbk}.
Farkas found zip bifurcations while studying the competition dynamics involving
one prey and two predator species:
\begin{eqnarray}
\dot S &=& \gamma S\Big(1-\frac{S}{K}\Big)
             -\frac{m_1x_1S}{a_1+S} - \frac{m_2x_2S}{a_2+S},\label{e1}\\
\dot x_1 &=& \frac{m_1x_1S}{a_1+S} - d_1x_1,\label{e2}\\
\dot x_2 &=& \frac{m_2x_2S}{a_2+S} - d_2x_2.\label{e3}
\end{eqnarray}
In these equations, $x_1, x_2, S$  denote the population size of the two
predator and the prey species, respectively.
Clearly, in the absence of predators, the prey follow a standard logistic
growth whose increase is controlled by $\gamma>0$.
The carrying capacity of the environment with respect to the prey is $K>0$.
The impact of the predators is assumed to be regulated by the
Michaelis-Menten kinetics \cite{mm}, where $m_i>0$, $d_i>0$, and $a_i$ are,
respectively, the maximum birth rate, the death rate and the half saturation
constant of the $i$-th predator.

Now, introduce auxiliary variables
\[ \lambda_i=\frac{a_id_i}{m_i-d_i}, \qquad \beta_i=m_i-d_i, \qquad
     b_i= m_i/d_i, \]
and assume that $0<\lambda=\lambda_1=\lambda_2 <K$, 
implying $\beta_i>0$, $b_i>1$.
These definitions change  Eqs.~(\ref{e1})-(\ref{e3}) into
\begin{eqnarray}
\dot S &=& \gamma S\Big(1-\frac{S}{K}\Big)
             -\frac{m_1x_1S}{a_1+S} - \frac{m_2x_2S}{a_2+S},\label{eq1}\\
\dot x_1 &=& \beta_1x_1\frac{S-\lambda}{a_1+S},\label{eq2}\\
\dot x_2 &=& \beta_2x_2\frac{S-\lambda}{a_2+S}.\label{eq3}  
\end{eqnarray}
Clearly, in the non-negative octant $\mathbb R_+^3$
the equilibria of the equations above are $(0,0,0)$, $(K,0,0)$ 
as well as the points on the straight line segment $L$ \cite{fbk,obit}:
\begin{equation}
 L =\Big\{ (\lambda,\xi_1,\xi_2) \in \mathbb R_+^3 \quad\Big\vert\quad
\frac{m_1\xi_1}{a_1+\lambda} +\frac{m_2\xi_2}{a_2+\lambda}
  = \gamma\big(1-\frac{\lambda}{K}\big)  \Big\} 
\end{equation}

To see what Farkas calls zip bifurcation, fix parameters as follows \cite{obit}:
\[ \gamma=\lambda=a_2=1, \qquad a_1=m_2=2, \qquad m_1=3, \]
and consider the triplet of points
\begin{eqnarray*}
 P_K&=&\Big(1,0,1-\frac{1}{K}\Big), \\
   M_K&=&\Big(1,3-\frac{9}{K},\frac{8}{K}-2\Big), \\
   Q_K&=&\Big(1,1-\frac{1}{K},0\Big),
\end{eqnarray*}
where $K\in (3,4)$.
With the choices above, it is not difficult to see that the straight line
segment $L$ connects the points $P_K$ and $Q_K$, and that the point $M_k$
is contained in the line $L$.
As discussed in Section 7.4 of Farkas' book \cite{fbk},
the equilibria on $L$  located between $P_K$ and $M_k$ are unstable,
while the equilibria between $M_k$ and $Q_K$ are stable.
As $K$ increases from $3$ to $4$, the point $M_k$ moves on the line from
$P_K$ to $Q_K$, so that all points located on the left of $M_k$ become
unstable. As $K$ is varied, the line $L$ undergoes a parallel displacement
which, however, has no effect on the aforementioned scenario. 
For additional references and examples in more general contexts and the
corresponding analysis, see Section 7.4 of Farkas' book \cite{fbk}.

More recent work by Ferreira and Rao deals with zip-bifurcation in a
predator-prey model with diffusion \cite{fo09}, and in systems involving
discrete delay \cite{fft,fr13} and cross-diffusion \cite{fsr19}.
Zip bifurcations are also discussed by Escobar-Callejas et al.~\cite{cal10},
and by Echeverri et al.~\cite{e17}.

\section{Velcro bifurcations}

Velcro bifurcations were considered in 2003 by Bocs\'o and Farkas \cite{bf},
in the context of a political and economic
rationality  economic problem modelled  by a set of four differential equations
taking into account information concerning the problem spread among the people
who support the political alternatives.
In such model, velcro bifurcation occurs for specific parameter combinations
destabilizing the equilibrium points when information spreads \cite{bf}.

The model consists of the following equations:
\begin{eqnarray}
\dot v &=& \gamma v\big(1-\frac{y}{K}\big)
  -\sum_{i=1}^3 m_i\frac{v}{a_i+v}u_i - M\frac{v}{A+v}\frac{u_1}{u_2},\\
\dot u_1 &=& m_1\frac{v}{v+a_1}u_1 - d_1u_1,\\
\dot u_2 &=& m_2\frac{v}{v+a_2}u_1 - d_2u_2,\\
\dot u_3 &=& m_3\frac{v}{v+a_3}u_1 - d_3u_3.
\end{eqnarray}
The model has similarities with the previous one, control parameters obey similar
relations but have rather different meanings \cite{bf} which are of no concern
for our purpose here.

The last three equations of the model above may be simplified to
\begin{equation}
  \dot u_i = \beta_i\frac{v-\lambda_i}{v+a_i}u_i, \qquad i=1,2,3
\end{equation}
where, similarly as before,
\[ \beta_i=m_i-d_i, \qquad b_i=m_i/d_i, \qquad
  \lambda_i=\frac{a_id_i}{\beta_i} = \frac{a_i}{b_1-1}.   \]
Under specific but realistic relations of the parameters the equilibrium
points of the system form a surface \cite{bf}
\begin{eqnarray*}
  S=\Big\{ &&(v,u_1,u_2,u_3) \in\mathbb R^4 \quad\vert\quad
     v=\lambda,u_1,u_2,u_3>0,\\
    &&  \gamma\Big(1-\frac{v}{K}\Big)=
  \sum_{i=1}^3\Big(\frac{m_iu_i}{a_i+v}+\frac{M}{A+v}\frac{u_1}{u_2}\Big)\quad \Big\}
\end{eqnarray*}

In the above context, the dynamics of the velcro bifurcations is defined
as a sort of generalized zip bifurcations, as spelled out in Theorem 2 of
Bocs\'o and Farkas \cite{bf}:
\begin{menor}{0.8truecm}
{\em The surface $S$ is divided into two parts by a curve $g$; the equilibria on
the upper part of $S$ are still stable, and this part is an attractor of
the system (in the sense described in \cite{fbk});
the equilibria on the lower parts are already unstable. The curve $g$
moves upwards as $K$ is increased leaving behind the destabilized equilibria.}
\end{menor}

Velcro bifurcations were also reported in 2005 by Farkas, S\'aez, and Sz\'ant\'o
in competition models with generalized Holling functional response \cite{fss,s06}.
Specifically, their basic model is given by the equations:
\begin{eqnarray}
\dot S &=& rS\big(1-\frac{S}{K}\big) -\sum_{i=1}^3m_ix_i\frac{S^n}{a_i^n+S^n},\\  
\dot x_i &=& m_ix_i \frac{S^n}{a_i^n+S^n} - d_i x_i,  \qquad i=1,2,3
\end{eqnarray}
where $n>2$ is an integer. For a detailed analysis and several figures of
the equilibrium surfaces, consult the original article \cite{fss,s06}.

\section{Conclusions and outlook}

The purpose of this note is to bring the works of Farkas and co-workers to the
attention of researchers working in the interesting field of systems with
line and surfaces of equilibria.
As remarked in 1996 by Freedman \cite{f96},
``For those who don't know, zip bifurcations were first discovered by
Professor Farkas, describing how a singular curve unfolds into periodic solutions
when a parameter changes, just like a zipper opening up.''

It is interesting to note that while most of the recent systems found to contain
lines and surfaces of equilibria deal with interesting but abstract
polynomial systems, arising from exhaustive computer searches, that are not yet
associated with any applications.
In contrast,  Farkas and co-workers found zip and velcro bifurcations in
standard systems that contain  typical nonlinearities of the sort encountered in
popular models used in biology and economy.
A publication list containing 76 works of Farkas is given in Ref.~\cite{obit},
while a list with 78 works, signed by ``students and colleagues'', is given
in Ref.~\cite{obit2}. A special issue of the journal
{\sl Differential Equations
and Dynamics Systems} was dedicated to Farkas \cite{si09}.  

As it is clear from the literature,
there is presently great interest in investigating changes in the topology of
attractors not restricted to small neighborhoods of points.
The works of Farkas still contain a plethora of theorems and unexplored materials
that deserve attention, and that will certainly contribute to the understanding
of the rich dynamics of systems with equilibria defined by extended
mathematical structures.

\bigskip\bigskip

The author is indebted to J.D.~Ferreira, S.~Jafari, and V.T.~Pham for their
interest, and the first one for pointing out Ref.~\cite{fsr19} to him.
Work done in the framework of an 
{\sl Advanced Study Group} on {\it Forecasting with Lyapunov vectors},
at the Max-Planck Institute for the Physics of Complex Systems, Dresden.
The author was  supported by CNPq, Brazil.


\end{document}